\documentclass{aa}
\begin{document}
\outer\def\gtae {$\buildrel {\lower3pt\hbox{$>$}} \over 
{\lower2pt\hbox{$\sim$}} $}
\outer\def\ltae {$\buildrel {\lower3pt\hbox{$<$}} \over 
{\lower2pt\hbox{$\sim$}} $}
\newcommand{\ergscm} {ergs s$^{-1}$ cm$^{-2}$}
\newcommand{\ergss} {ergs s$^{-1}$}
\newcommand{\ergsd} {ergs s$^{-1}$ $d^{2}_{100}$}
\newcommand{\pcmsq} {cm$^{-2}$}
\newcommand{\ros} {\sl ROSAT}
\newcommand{\exo} {\sl EXOSAT}
\def\rchi{{${\chi}_{\nu}^{2}$}}
\newcommand{\Msun} {$M_{\odot}$}
\newcommand{\Mwd} {$M_{wd}$}
\def\Mdot{\hbox{$\dot M$}}
\newcommand{\dg} {$^{\circ}$}

\def\cs{~count~s$^{-1}$}        
\def\cks{~count~ks$^{-1}$}     
\def\cpix{~count~pix$^{-1}$}     
\def\x{$\times$}
\def\mum{~${\rm \mu m}$}    
\def\exp{\times 10^}

%

   \title{The XMM-Newton Optical/UV Monitor Telescope}

   \author{K. O. Mason\inst{1}, A. Breeveld\inst{1}, R. Much\inst{2},
	M. Carter\inst{1}, F. A. Cordova\inst{3}, M. S. Cropper\inst{1}, 
	J. Fordham\inst{4}, H. Huckle\inst{1}, C. Ho\inst{6},
	H. Kawakami\inst{1}, J. Kennea\inst{3}, T. Kennedy\inst{1}, 
	J. Mittaz\inst{1}, D. Pandel\inst{3},
	W. C. Priedhorsky\inst{6}, T. Sasseen\inst{3}, R. Shirey\inst{3},
	P. Smith\inst{1}, J.-M. Vreux\inst{5}}

   \offprints{Keith Mason: kom@mssl.ucl.ac.uk}

\institute{$^{1}$Mullard Space Science Laboratory, University College London,
Holmbury St. Mary, Dorking, Surrey, RH5 6NT, UK\\
$^{2}$Astrophysics Division, Space Science Department of ESA, P.O. Box 299,
NL-2200 AG Noordwijk, Netherlands\\
$^{3}$Department of Physics, University of California, Santa
Barbara, California 93106, USA\\
$^{4}$Department of Physics and Astronomy, University College London, 
Gower Street, London\\
$^{5}$Institut d'Astrophysique et de G\'{e}ophysique,
Universit\'{e} de Li\`{e}ge,
5, Avenue de Cointe,
B-4000-Li\`{e}ge,
Belgium\\
$^{6}$Los Alamos National Laboratory,
           PO Box 1663,
           Los Alamos, NM 87545
           USA}

\authorrunning{K. O. Mason  et~al.~}
\titlerunning{XMM-OM Telescope}

\date{}

\maketitle

\begin{abstract}

The XMM-OM instrument extends the spectral coverage of the XMM-Newton observatory
into the ultraviolet and optical range. It provides imaging and time-resolved 
data on targets simultaneously with observations in the EPIC and RGS. It also
has the ability to track stars in its field of view, thus providing an improved
post-facto aspect solution for the spacecraft. An overview of the XMM-OM and
its operation is given, together with current information on the performance of
the instrument. 


\end{abstract}

\section{Introduction}

The Optical/UV Monitor Telescope (XMM-OM) is a standalone instrument
that is mounted on the mirror support platform of XMM-Newton
(Jansen et~al.~2001) alongside
the X-ray mirror modules. It provides coverage between 170~nm and 650~nm
of the central 17 arc minute square region of the X-ray field of view (FOV),
permitting routine multiwavelength observations of XMM targets
simultaneously in the X-ray and ultraviolet/optical bands. Because of
the low sky background in space, XMM-OM is able to achieve impressive
imaging sensitivity compared to a similar instrument on the ground,
and can detect a $B=23.5$ magnitude A-type star in a 1000~s integration in
``white'' light (6 sigma).
It is equipped with a set of broadband filters for colour discrimination.
The instrument also has grisms for low-resolution spectroscopy, and an image
expander (Magnifier) for improved spatial resolution of sources.  Fast
timing data can be obtained on sources of interest simultaneously with image
data over a larger field.

In the following sections we give an overview of the instrument
followed by an account of its operation in orbit and the instrument 
characteristics.

\section{Instrument overview}

The XMM-OM consists of a Telescope Module and a separate Digital
Electronics Module, of which there are two identical units for
redundancy (see Fig. 1).  The Telescope Module
contains the telescope optics and detectors, the detector processing
electronics and power supply.  There are two distinct detector chains,
again for redundancy. The Digital Electronics Module houses the
Instrument Control Unit, which handles communications with the
spacecraft and commanding of the instrument, and the Data Processing
Unit, which pre-processes the data from the instrument before it
is telemetered to the ground.

\begin{figure*}
\begin{center}
\setlength{\unitlength}{1cm}
\begin{picture}(8.8,10)
\put(10,0){\includegraphics{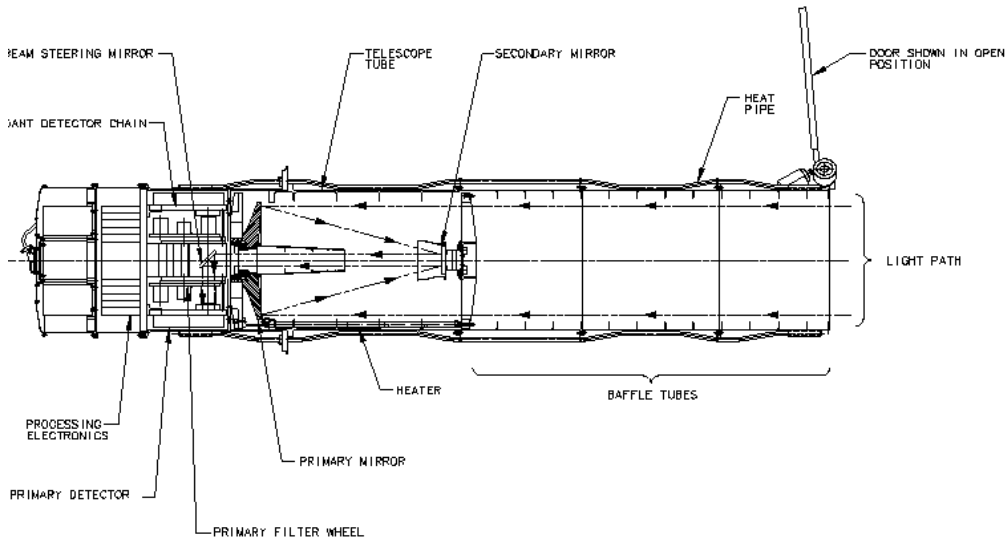}}
\end{picture}
\caption{A mechanical drawing of the XMM-OM telescope module showing the 
light path through to the detectors.}
\label{figure1}
\end{center} 
\end{figure*}

\subsection{Optics} 

The XMM-OM uses a Ritchey Chr\'{e}tien telescope design modified by
field flattening optics built into the detector window. The f/2 primary
mirror has a 0.3~m diameter and feeds a hyperboloid secondary which modifies
the f-ratio to 12.7. A $45$\dg\ flat mirror located behind the primary can
be rotated to address one of the two redundant detector chains. In each chain
there is a filter wheel and detector system. The filter wheel has 11
apertures, one of which is blanked off to serve as a shutter, preventing
light from reaching the detector. Another seven filter locations house 
lenticular filters, six of which constitute a set of broad band filters for
colour discrimination in the UV and optical between 180~nm and 580~nm 
(see Table 2 for a list of filters and their wavelength bands). The
seventh is a ``white light'' filter which transmits light over the full range
of the detector to give maximum sensitivity to point sources. The 
remaining filter positions contain two grisms, one optimised for the 
UV and the other for the optical range, and a \x 4 field expander 
(or Magnifier) to provide
high spatial resolution in a 380--650~nm band of the central portion of the (FOV).

\subsection{Detector} 

The detector is a microchannelplate-intensified CCD (Fordham et~al.~1992). 
Incoming photons are converted into photoelectrons in an S20 photocathode
deposited on the inside of the detector window. The
photoelectrons are proximity focussed onto a microchannelplate
stack, which amplifies the signal by a factor of a million,
before the resulting electrons are converted back into photons by
a P46 phosphor screen. Light from the phosphor screen is passed
through a fibre taper which compensates for the difference in
physical size between the microchannelplate stack and the fast-scan 
CCD used to detect the photons. The resulting photon splash on the CCD covers
several neighbouring CCD pixels (with a FWHM of approximately 
1.1 CCD pixels, if fitted with a Gaussian). 
The splash is centroided, using a 3\x 3 CCD pixel subarray
to yield the position of the incoming photon to a fraction of a CCD
pixel (Kawakami et~al.~1994). An active area
of 256\x 256 CCD pixels is used, and incoming photon events are
centroided to 1/8th of a CCD pixel to yield 2048\x 2048 pixels
on the sky, each 0.4765 arc seconds square. 
In this paper, to avoid confusion, while CCD pixels 
(256\x 256 in FOV) will be referred to explicitly, a pixel refers 
to a centroided pixel (2048\x 2048 in FOV). 
As described later, images are normally taken 
with pixels binned 2\x 2 or at full sampling. 

The CCD is read out
rapidly (every 11~ms if the full CCD format is being used) to
maximise the coincidence threshold (see sect. 5.2).

\subsection{Telescope mechanical configuration}  

The XMM-OM telescope module consists of
a stray light baffle and a primary and secondary mirror assembly,
followed by the detector module, detector processing electronics and telescope
module power supply unit. The separation of the primary and secondary
mirrors is critical to achieving the image quality of the telescope. The
separation is maintained to a level of 2\mum\ by invar support rods
that connect the secondary spider to the primary mirror mount. Heat
generated by the detector electronics is transferred to the baffle by heat
pipes spaced azimuthally around the telescope, and radiated into space. In
this way the telescope module is maintained in an isothermal condition, at a
similar temperature to the mirror support platform. This minimizes changes in
the primary/secondary mirror separation due to thermal stresses in the invar
rods. Fine focussing of the telescope is achieved through two sets of
commandable heaters. One set of heaters is mounted on the invar support
rods. When these heaters are activated, they cause the rods to expand,
separating the primary and secondary mirrors. A second set of heaters on the
secondary mirror support brings the secondary mirror closer to the primary
when activated. The total range of fine focus adjustment available is 
$\pm10\mu$m.

The filter wheel is powered by stepper motor, which drives the wheel in one
direction only. The filters are arranged
taking into account the need to distribute the more massive elements
(grisms, Magnifier) uniformly across the wheel.

\subsection{Digital Electronics Module} 

There are two identical Digital Electronics 
Modules (DEM) serving respectively the two redundant detector chains.
These units are mounted on the mirror support platform, separate from the
telescope module. Each DEM contains an Instrument Control Unit (ICU) and a
Digital Processing Unit (DPU). The ICU commands the XMM-OM and handles
communications between the XMM-OM and the spacecraft.

The DPU is an image processing computer that 
digests the raw data from the
instrument and applies a non-destructive compression algorithm before
the data are telemetered to the ground via the ICU. The DPU supports two main
science data collection modes, which can be used simultaneously. 
In Fast Mode, data from a small region of the
detector are assembled into time bins. 
In Image Mode, data from a large region are extracted to create an
image. These modes are described in more detail in the
next sect.
The DPU autonomously selects up to
10 guide stars from the full XMM-OM image and monitors their position in
detector coordinates at intervals that are typically set in the range 10--20
seconds, referred to as a tracking frame. These data provide a record of the
drift of the spacecraft during the observation accurate to $\sim 0.1$
arc second. The drift data are used within the DPU to correct Image Mode
data for spacecraft drift (see sect. 5.5).

\section{Observing with XMM-OM}

\subsection{Specifying windows}

The full FOV of XMM-OM is a square 17\x 17 arc minutes,
covering the central portion of the X-ray FOV. Within this field
the observer can define a number of data collection windows around targets
or fields of interest. Up to five different Science Windows can be defined
with the restriction that their boundaries may not overlap. However, one
window can be completely contained within another.

Because of constraints on the telemetry rate available, it is not possible
to transmit the full data on every photon that XMM-OM detects.
Instead a choice has to be made between image coverage and time resolution.
Thus two types of Science Window can be defined, referred to as Image Mode
and Fast Mode. A maximum of two of the five available science windows
can be Fast Mode.

\it Image Mode \rm emphasizes spatial coverage at the expense of timing
information. Images can be taken at the full sampling of the instrument
or binned by a factor of 2 or 4, to yield a
resolution element on the sky of approximately 0.5, 1.0 or 2.0 arc seconds 
(a factor of four finer for the Magnifier). 
The maximum total size of the Science Windows is determined by the memory
available in the DPU. A single Image Mode window binned by a factor of
2\x 2 
can be up to 976\x 960 detector pixels, which
results in a 488\x 480 binned
pixel image being stored in the DPU. At full sampling (with no binning) the window can be up 
to 652\x 652 pixels.
Any drift in the pointing direction of the spacecraft is corrected in the
image by tracking guide stars (section 5.5).

\it Fast Mode \rm emphasizes timing information at the expense of spatial coverage.
The maximum total number of pixels that can be specified for a Fast Mode
window is 512. Thus the maximum size of an approximately square window would
be 22\x 23 pixels ( = 506 total). Note that there is no binning within a Fast
Mode window. The pixel locations of individual photons within the window
are recorded and assigned a time tag, which has a user-specified integration 
time of between 100~ms and the tracking frame duration (10--20~s). 
No tracking correction is applied to Fast Mode data. This can be applied
on the ground, from the drift history supplied by XMM-OM.

To simplify observation set-up, two standard observing sequences of five
exposures have been created that together cover the whole XMM-OM FOV
at one arc second sampling while at the same time monitoring a
central target at full spatial sampling (0.5 arcsec). In the first variant,
each of the five exposures contains an unbinned Image Mode window centred on
the prime instrument's boresight (the position of the main target), 
and a second Image Mode window, binned by 2\x 2
pixels, that is defined in each of the set of five exposures so as to
form a mosaic of the entire field (see Fig. 2). 
The second variant is exactly the same as the first except that a Fast Mode
window is added around the prime instrument's boresight position.

\begin{figure}
\begin{center}
\leavevmode
\setlength{\unitlength}{1.0cm}
\begin{picture}(8.8,8)
\put(0,0){\includegraphics{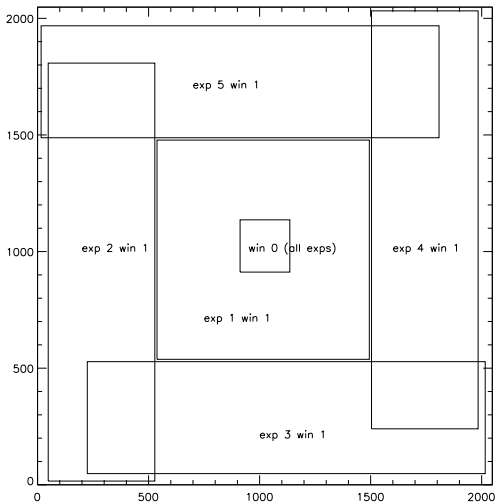}}
\end{picture}
\caption{Setup of XMM-OM imaging default mode. There are
5 exposures, each one being made up of two image mode windows: win 0 and
win 1. Win 0 is the same for each exposure and is unbinned
(0.5 arcsecs). Win 1 for each exposure is binned 2\x 2 pixels 
and the position changes for each exposure.
For exposure 1, win 1 is a square containing the central half of
the FOV, with win 0 entirely inside it. For the other exposures,
win 1 is arranged to fill in the rest of the FOV.}
\label{figure2} 
\end{center}
\end{figure}

The length of an XMM-OM Image Mode only exposure can be set in the range
800--5000~s. However it should be noted that there is an approximately 300
second overhead associated with each individual exposure. The maximum length 
of an exposure that contains a single Fast Mode window is 4400~s, or 2200~s
if there are two Fast Mode windows.

At the time of writing a further mode is being commissioned which allows the
full field to be imaged at 1 arcsec sampling in one go, at the expense of
tracking information and correction. This is made possible by the impressive
stability of the XMM-Newton spacecraft compared to pre-launch expectation.

Window coordinates can be specified either in detector pixels, or in sky
coordinates. To facilitate the latter, the XMM-OM performs a short V-band
observation at the start of each pointing. The DPU compares the image with
the positions of uploaded field stars to calibrate
the absolute pointing of the OM.

\subsection{Filter selection}

The XMM-OM filter wheel rotates in one direction only and, to conserve the
total number of wheel rotations over the expected lifetime of XMM-Newton,
the number of filter wheel rotations per pointing is limited to one (unless 
there are very strong scientific arguments for more). Thus
filter observations have to be executed in a particular order during a given
target pointing. The filter elements are listed in the order
they occur in the filter wheel in Table 2. 
The instrument
is slewed with the blocked filter in place, and thereafter a field
acquisition exposure is performed in V.

\begin{table}
\caption{OM filters, with the V-magnitude brightness limits 
for the main optical and UV filters, for a range of stellar types,
assuming that they are not in the centre of the FOV. (Note the order
of the filter elements in this table is not the order in which they should
be used; see table 2)}
\label{table1} 
\halign{#\hfil\quad & #\hfil\quad &
		\hfil#\hfil\quad &
		\hfil#\hfil\quad &
		\hfil#\hfil\quad &
		\hfil#\hfil\quad &
		\hfil#\hfil\cr
            &  B0   & A0   & F0    & G0    & K0    & M0\cr
V           & 7.71  & 7.68  & 7.65  & 7.65  & 7.63  & 7.59\cr
B           & 9.38  & 9.18  & 8.83  & 8.58  & 8.29  & 7.68\cr
U           & 9.79  & 8.34  & 7.88  & 7.57  & 6.66  & 4.50\cr
UVW1        & 9.49  & 7.55  & 6.53  & 5.98  & 3.89  & 1.50\cr
UVM2        & 8.94  & 6.82  & 4.53  & 2.70 & -0.23 & -2.63\cr
UVW2        & 8.76  & 6.55  & 3.83  & 1.86 & -0.63 & -1.80\cr
Mag         & 10.03 & 9.68  & 9.41  & 9.27  & 9.13  & 8.91\cr
White       & 11.58 & 10.28 & 9.72  & 9.50  & 9.22  & 8.93\cr}
\end{table}
 
The same telescope focus setting is used for all the filters except for the
Magnifier (see sect. 5.6.1), 
where the optimum focus is different (the image quality is the
most sensitive to focus position when using the Magnifier).
 
The XMM-OM instrument is optimised for the detection of faint sources. If
the source count-rate is too high the response of the detector is
non-linear. This ``coincidence loss'' occurs when the probability of 
more than one photon splash
being detected on a given CCD pixel within the same CCD readout frame
becomes significant. Coincidence loss is discussed in more detail in
sect. 5.2. If a source is predicted to exceed the
coincidence threshold for a given filter, then a different filter with lower
throughput can be selected. Alternatively a grism can be
selected which disperses the available light over many pixels.

The XMM-OM detectors can also be permanently damaged by exposure to a source
that is too bright, reducing both the quantum efficiency of the photocathode
and the gain of the channelplates. This is a cumulative effect dependent on
the total number of photons seen over the lifetime of the instrument at a
particular location on the detector. The deterioration is therefore more
severe for longer observations of a bright source.  
For this reason limits are imposed on
the maximum brightness of stars in the FOV (see Table 1) and apply to any 
star in the FOV irrespective of whether it is within a science window or
not. Even more stringent limits are applied to the central region of the
detector that will usually contain the target of interest. In the event that
there is a star in (or near to) the FOV that violates the
brightness constraints, a different filter, which has lower photon
throughput, can be selected. Also, if the bandpass is
appropriate, the Magnifier
can be used to exclude bright stars further than a few arc minutes from the
field centre.

The grisms (one optimised for the UV, the other for the optical) form a
dispersed first order image on the detector, together with a zeroth order
image that is displaced in the dispersion direction. The counts in the
zeroth order image of field stars determines the brightness limits used for
observing with the grisms.

\section{OM performance}

\begin{figure*}
\begin{center}
\setlength{\unitlength}{1cm}
\begin{picture}(8.8,10)
\put(0,10){\includegraphics{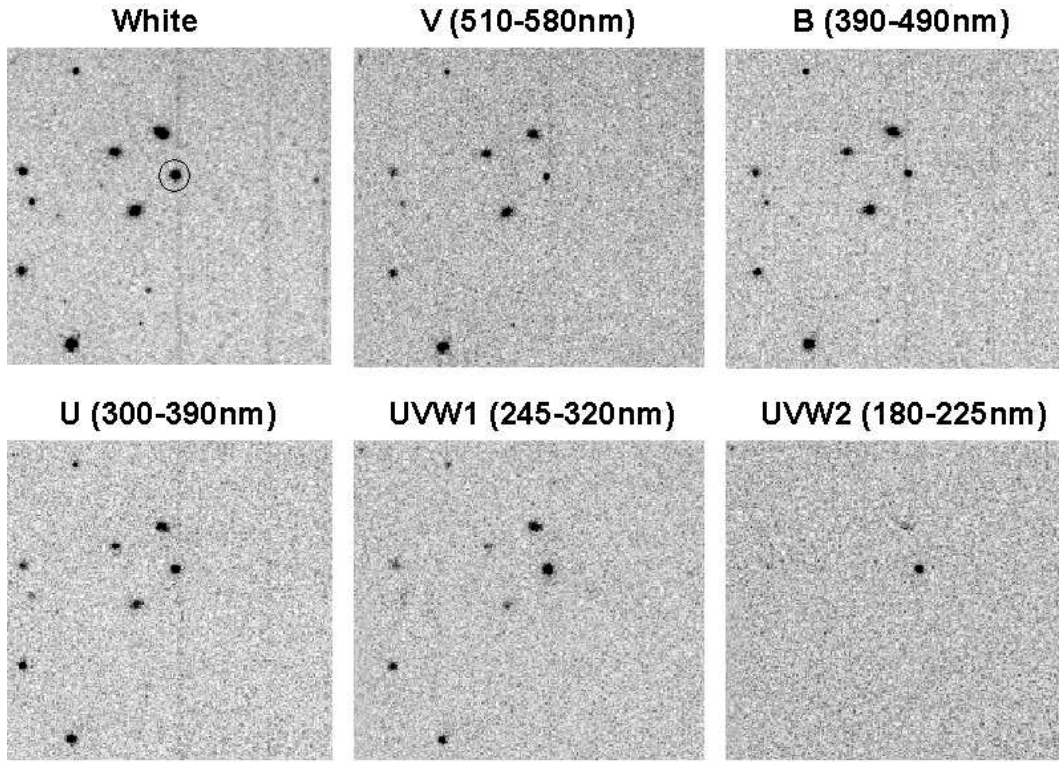}}
\end{picture}
\caption{An image of part of the Lockman Hole field 
taken in various XMM-OM filters
as marked. The object circled in the White Light image is an AGN, referred
to as R32 by Schmidt et~al.~(1998). The size of the images is about 3.5 arc
minutes across and the exposure times were 1500~s (white), 1000~s (V), 1000~s
(B), 1000~s (U), 2200~s (UVW1) \& 4400~s (UVW2). Faint vertical bars in some of
the images are caused by charge leakage along the readout direction of  the
CCD from bright stars outside the illustrated field.}
\label{figure3}
\end{center} 
\end{figure*}

The first light observation for XMM-OM took place on 2000 January 11. Since
then the various engineering and science data taking modes of the instrument
have been commissioned including full-field image engineering mode (not
generally available for science observations because of the very large
telemetry overhead required to transmit the data to the ground), and the
Image and Fast science modes. The telescope focus has been optimised using
the heater-based fine focus control, and the gain of the image intensifier
has been optimised.  The performance of the DPU in tracking image motion due
to spacecraft drift has also been verified and distortion maps derived to
relate XMM-OM detector coordinates to the sky. Photometric calibrations have
been derived for all filter elements, but work continues on colour equations
and to tie these more accurately into standard systems.  Similarly,
preliminary throughput and wavelength calibrations have been derived for the
Grisms.

To illustrate the capabilities of XMM-OM, we show in Figure 3 images of
part of the Lockman Hole field in the White Light filter, and in five of the
six colour filters (the remaining filter, UVM2, was not used during this
observation). The images contains an $R=18.1$ magnitude AGN identified in the
ROSAT observation of the field, and referred to as R32 by Schmidt et~al.~
(1998). The AGN is clearly UV bright. The XMM-OM detects approximately 12\cs\
from the AGN in White Light, while the count rate in the
colour filters ranges from a high of 2.9\cs\ in U, down to about 0.25\cs\ in
the UV filter UVW2.

To illustrate the spectral capability of XMM-OM, we show in Figure 4 the
extracted spectrum of the DA white dwarf standard BPM16274. The Balmer
absorption lines can be clearly discerned.

\begin{figure}
\begin{center}
\setlength{\unitlength}{1cm}
\begin{picture}(8.8,7)
\put(0,0){\includegraphics{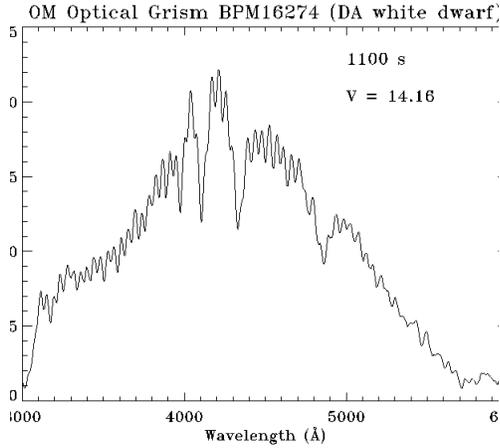}}
\end{picture}
\caption{Spectrum of the white dwarf BPM16274 obtained using the visual grism. The 
regular low-level fluctuations in the signal with wavelength are due to the 
modulo-8 pattern (section 5.6.3), which has not been corrected in this plot. 
}
\label{figure4}
\end{center} 
\end{figure}

\section{Analysis issues}

\subsection{Throughput}

An initial estimate of the zero points of the various 
XMM-OM broadband filters (i.e. the magnitude which yields 1 count per second; Table 2.)
was derived from calibration observations of two white dwarfs.
\begin{itemize}
\item The OM response throughput was determined based on  measurements of 
      the spectrophotometric standards BPM16274 and LBB227. 
\item Using the OM throughput model  an OM exposure of Vega was simulated.
\item The zeropoints in U-, B-, V-filter were fixed in a way that the 
      brightness of  Vega matches the literature values. 
\item In the UV-filters the brightness of Vega was set to 
       0.025 mag; the average {\it U} magnitude of Vega in the literature being
       the most appropriate for the UV filters, 
       and to 0.03 mag (the average Vega {\it V} magnitude) for the broadband 
       Magnifier and the White Light filters.
\end{itemize}

The calculated OM zeropoints are written into the relevant XMM-CCF file, 
to be used by the XMM Science Analysis System (SAS) (Watson et~al.~2001).
Updates of the zeropoint definitions as well as more
precise colour transformations (Royer et~al.~2000) to the standard UBV system
are expected once the results of a
dedicated ground based photometric observation programme become available.
In the framework of this programme several OM calibration fields are
anchored to high quality secondary photometric standards deep fields
established by ESO.
Early results of this ground observations are expected in October 2000.

\begin{table}
\caption{The OM filters, their wavelength
bands in nm, and the preliminary zero points. The zero points will be  
updated once the data of the ground based observation programme 
become available in October 2000. The filters are in
the order they occur in the filter wheel.}
\label{table2} 
\halign{#\hfil\quad &
		\hfil#\hfil\quad &
		\hfil#\hfil\cr
Filter          &  wavelength band (nm) & zeropoints (mag)\cr   
Blocked\cr
V               & 510--580 &  18.11\cr         
Magnifier       & 380--650\cr         
U 		& 300--390 &  18.24\cr         
B 		& 390--490 &  19.28\cr         
White (clear)	& 150--500 \cr  
Grism 2 (vis) 	& 290--500\cr
UVW1 		& 245--320 &  17.37\cr
UVM2 		& 205--245 &  16.02\cr      
UVW2 		& 180--225 &  15.17\cr      
Grism 1 (UV)	& 200--350\cr}
\end{table}
 
From analysis of the Lockman Hole field, avoiding the central 1 arc minute
of the FOV where the background is enhanced (section 5.6.4)
the limiting magnitude after 1000~s is calculated to be 21.0 in {\it V},
22.0 in {\it B} and 21.5 in {\it U} (6 sigma). 
The limiting magnitude for the White Light filter is very dependant on
the spectral type of the star, because the bandpass is so
broad. However, 
for an A0 star we estimate that the 6 sigma limiting magnitude above background is
$\sim$23.5.

\subsection{Coincidence loss and deadtime}

Coincidence loss is observed whenever the count rate is such that more than
one photon arrives in the same place within a given readout frame. 
Losses become significant for a point source at a
count rate of about $10$\cs\ (for 10\% coincidence) 
when the full CCD chip is
being readout (i.e. about 2.5 magnitudes brighter than the zero points listed in Table 2). A factor of approximately two improvement can be
achieved by restricting the area of the CCD used, since this reduces the
time required to readout the chip.

The coincidence loss can be approximated by 

\begin{equation}
  ph_{\rm in}= {            \log{(1-{cts}_{\rm detected}*T)}  \over
   T_{\rm ft} - T }
\label{equation1}
\end{equation}

where\\
\begin{tabular}{l l}
$ph_{\rm in}$                & infalling photon rate per second \\
${cts}_{\rm detected}$   & measured count  rate per second \\
$T$           &CCD frametime in units of seconds \\
${T}_{\rm ft}$           & frametransfer time in units of seconds\\
\end{tabular}

Equation 1 applies strictly in the case of a perfectly point-like source. 
In practice a real stellar profile has wings, and the formula will 
break down at very high rates when coincidence among photons in the wings of
the profile becomes significant.  

The CCD deadtime depends on the size and shape of the science window used
but can be calculated accurately. The deadtime correction should be
applied by the SAS after any coincidence loss corrections.

\subsection{Flat fields}

An LED can be used to illuminate the detector by backscatter of
the photons from the blocked filter. These images are not completely flat
due to the illumination pattern of the LED, the gross shape of which could 
be removed by comparing with sky flats. However, using the LED allows a
large number of events to be collected in every pixel to give sufficiently 
high statistics for pixel to pixel sensitivity to be measured
and the relative measurement of any variation of the 
detector response on a fine scale. The LED brightness is adjustable and is
currently operated at a level that produces $3.25\exp{-3}$ \cs\ per binned (2\x
2) pixel.
So far, flat fields have been obtained to the level of 400 counts 
per binned (2\x 2) pixel allowing an accuracy of 5\% in the 
sensitivity measurement.
A CCF file in the SAS currently represents the accuracy of flat fields 
obtained before mid June, which is at the 10\% level.
Once sufficient flat fields have been obtained for a 2--3\% sensitivity
the relevant CCF file will be updated.

\subsection{Background}

The background count rate in the OM is dominated by the zodiacal light
in the optical. In the far UV the intrinsic detector background becomes
important. Images are regularly taken with the blocked filter and no
LED illumination to measure the detector dark counts.

The mean OM dark count rate is $2.56\exp{-4}$ \cs\ per pixel.  The variation in
dark count rate across the detector is $\pm 9\%$ and shows mainly a radial
dependence, being highest in an annulus at about 8 arcmin radius and lowest
in the centre.  When the
spacecraft is pointing at a very bright star, the dark rate is noticeably
increased (e.g.~up to 65\% higher for the $V=0$ star Capella) despite the blocked filter.
Excluding those dark frames taken during Capella
($V=0$) and Zeta Puppis ($V=2$) observations, the counts per dark frame
vary by only $\pm 7\%$ and show no trend of change with time.
  
\subsection{Tracking performance}

The positions of selected guide stars in the
XMM-OM FOV are measured each 10--20 second tracking frame, and an
X-Y offset applied to image mode data obtained during the tracking frame
before they are added to the master image in the DPU memory. The tracking
offsets are computed in pixels irrespective of the binning parameter
chosen. Using this ``Shift and Add'' technique, the final image 
is corrected on timescales greater than a few tens of
seconds and on spatial scales down to $\sim 0.5$ pixels, 
for drift in the pointing direction of the spacecraft.

The performance of tracking can be verified by comparing the PSFs of
stars taken during Fast Mode (at high time resolution and with no 
tracking) with those data
taken using Image Mode when tracking is enabled. 
This analysis has shown that XMM-OM tracking is performing as expected.

Analysis of OM tracking histories show that the spacecraft drift is less
than 0.5 pixels for approximately 75\% of all frames taken, and therefore
require no shift and add correction (a shift of one pixel will be made if
the guide stars are calculated to have drifted more than 
$\pm 0.5$ pixels from their reference positions). 
Of the remaining, the corrections due to drift are rarely
more than 2 pixels in any one direction.
 
Tracking is turned off automatically when no suitable guide stars are
found, which is usually due to poor statistics. 
This can occur in observations of very sparse
fields (rare) or when using the UVM2 and UVW2 filters, where throughput is 
lower than in the optical bands. However, given the pointing stability of
XMM-Newton and the intrinsically poorer resolution of the detector in the UV (section 5.6), 
this does not normally lead to any significant degradation 
in the PSF for non-magnified data.

\subsection{Image quality}

\subsubsection{Point Spread Function}

After launch the measured PSFs in the V-filter had FWHM widths
broader than 
expected from preflight measurements. The focus was therefore
adjusted using the control heaters as discussed in section 2.3. Fig. 5
shows the gradual change in the PSF with the heater setting.
As can be seen from the figure, the optimum setting for the
Magnifier is clearly at $-100$\% i.e. at the minimum separation
of the primary and secondary mirrors, whereas for the V-filter it is
above 70\%. A value of 100\% (maximum separation of mirrors) 
was chosen for subsequent measurements
in all filters, except for the Magnifier where $-100$\% is
selected. To allow for the
thermal settling time involved in a change of focus, twenty minutes of
additional overhead time is inserted before and after a sequence of
Magnifier exposures.

\begin{figure}
\begin{center}
\leavevmode
\setlength{\unitlength}{1.0cm}
\begin{picture}(8.8,5.5)
\put(0,0){\includegraphics{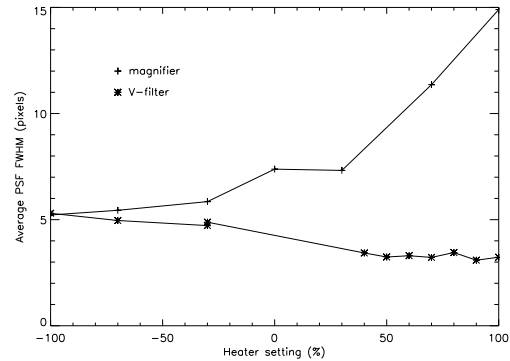}}
\end{picture}
\caption{PSF width with changing heater setting.
Each point in the V-filter represents the average of 23 stars in the
BPM16274 field}
\label{figure5} 
\end{center}
\end{figure}

The PSFs contain a contribution from the telescope optics and
from the detector. They can be assumed to be radially symmetric in shape, 
with an approximately gaussian central peak and extended
wings. The width of the PSF increases with photon energy because of the
detector component, from 3.1 pixels (1.5 arcsecs)
FWHM in the V band to $\sim 6$ pixels (3 arcsecs) 
pixels in the UV filters (see Fig. 6).

\begin{figure}
\begin{center}
\leavevmode
\setlength{\unitlength}{1.0cm}
\begin{picture}(8.8,6)
\put(0,0){\includegraphics{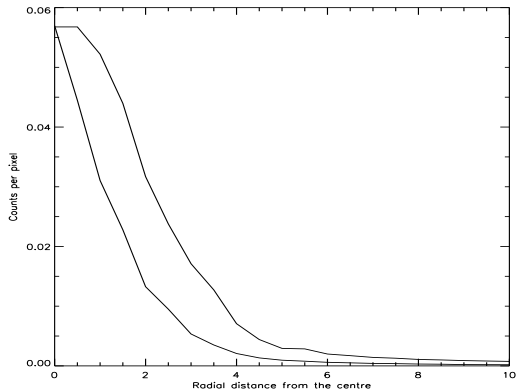}}
\end{picture}
\caption{Point spread function radial distribution. The inner curve is
the average PSF of 23 stars from the V-filter; the outer curve from
the UVW1-filter.}
\label{figure6} 
\end{center}
\end{figure}

\subsubsection{Distortion}

The XMM-OM optics, filters and (primarily) 
the detector system result in a certain amount
of image distortion. It is mainly in the form of barrel distortion, and if
not corrected
can result in shifts from the expected position of up to 20 arcsecs.
By comparing the expected position with the measured position for a large
number of stars in the FOV a distortion map has been derived.
The preliminary V-filter analysis was performed on the LMC pointing 
and is based
on 230 sources. A 3rd order polynomial was fitted to the deviations
assuming that there is no error at the centre of the FOV (i.e.
at address (1024.5,1024.5)). This polynomial can be
used to correct source positions measured in other fields, and currently gives
a positional RMS accuracy of 1.0 arcsec (1.9 pixels) in the V-filter
(see Fig. 7; astrometry relative to stars of known position over restricted
regions of the field can of course be more accurate than this).
Using higher orders of the polynomial does not increase the accuracy
and is detrimental particularly for sources at the edges of the FOV. 
Using functions other than polynomials has not yet been investigated,
but may lead to an improvement to the correction for sources near
the edges of the FOV.
Distortion maps using the 3C273 field have been derived
for the other filters, but are not yet to such high accuracy.
Further work will either use fields with more sources in the FOV
or combine data from several observations. 
The preliminary distortion maps have been entered into the 
appropriate CCF files
and can be used in conjunction with the SAS. They are also used
on board to automatically position windows on the detector that
are specified in sky coordinates. This is important for small
windows such as those used in Fast Mode.

\begin{figure}
\begin{center}
\leavevmode
\setlength{\unitlength}{1.0cm}
\begin{picture}(8.8,6)
\put(0,0){\includegraphics{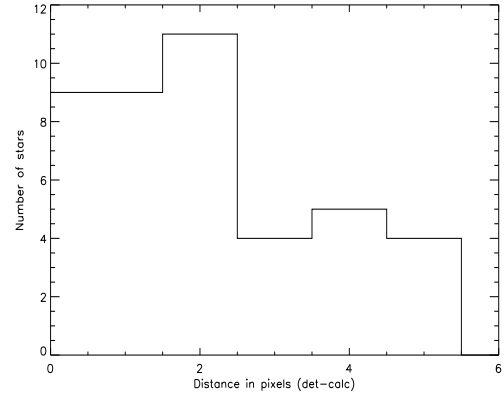}}
\end{picture}
\caption{Positional error of sources after the preliminary 
distortion correction.
This histogram was made using sources from the 3C273 field fitted
to a map derived from the LMC field. The higher deviations occur
near the edge of the FOV. The RMS positional deviation is 1.9 pixels,
equivalent to 1 arcsec.}
\label{figure7} 
\end{center}
\end{figure}

\subsubsection{Modulo-8 pattern} 

As discussed earlier, the XMM-OM detector functions by
centroiding a photon splash to within a fraction (1/8th) of a physical CCD
pixel. This calculation is performed in real-time by the detector
electronics, and therefore has to be fast. It is done by means of a lookup
table whose parameters are computed onboard once per revolution, 
based on a short image taken
with the internal flood LED lamp, and periodically updated. The lookup table
parameters are the mean values derived from a selected part of the active
area of the detector (usually the central region). They do not take into
account small variations in the shape of the photon splash over the 
detector face and as such are an
approximation to the optimum value at a given location on the detector.  

The result of imperfections in the lookup table is that the size of the
pixels is not equal on the sky. When displayed with a normal image
display routine, therefore, uncorrected XMM-OM images can exhibit a faint
modulation in the apparent background level repeating every eight
pixels, corresponding to every physical CCD pixel (see Fig. 8). SAS 
tasks that, for example, search for sources in XMM-OM images
take the variation in pixel size into account and compute the local 8\x 8
pattern post facto based on the measured image. Similarly the raw image can
be resampled for display purposes. The SAS routine does not lose or gain
counts, but resamples them according to the true pixel sizes.

The detector centroiding process also breaks down if more than one photon
splash overlaps on a given CCD frame. Thus an 8\x 8 pattern is often seen
around bright stars (see Fig. 8a), or when two bright stars occur close together on an
image.

\subsubsection{Scattered light} 

\begin{figure}
\begin{center}
\leavevmode
\setlength{\unitlength}{1.0cm}
\begin{picture}(10,12.0)
\put(0,0){\includegraphics{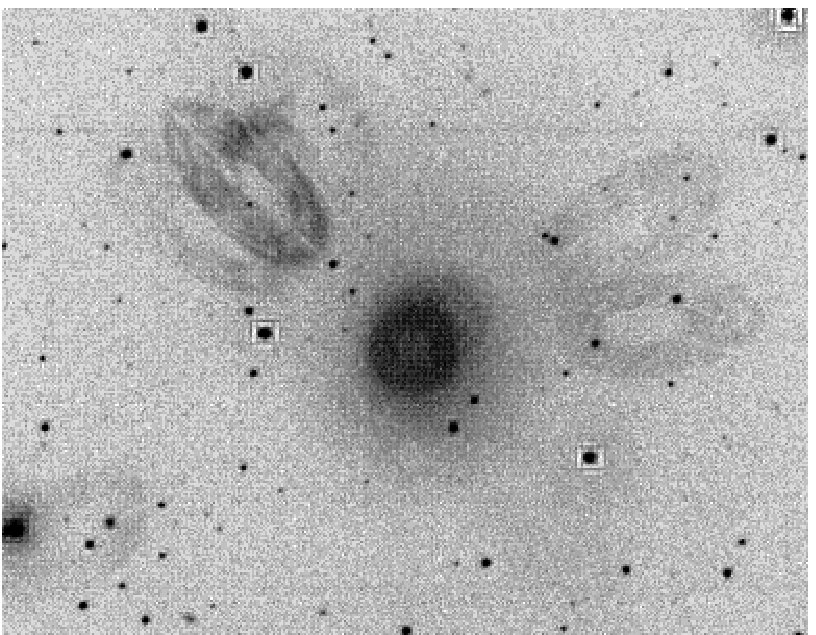}}
\put(0,6.5){\includegraphics{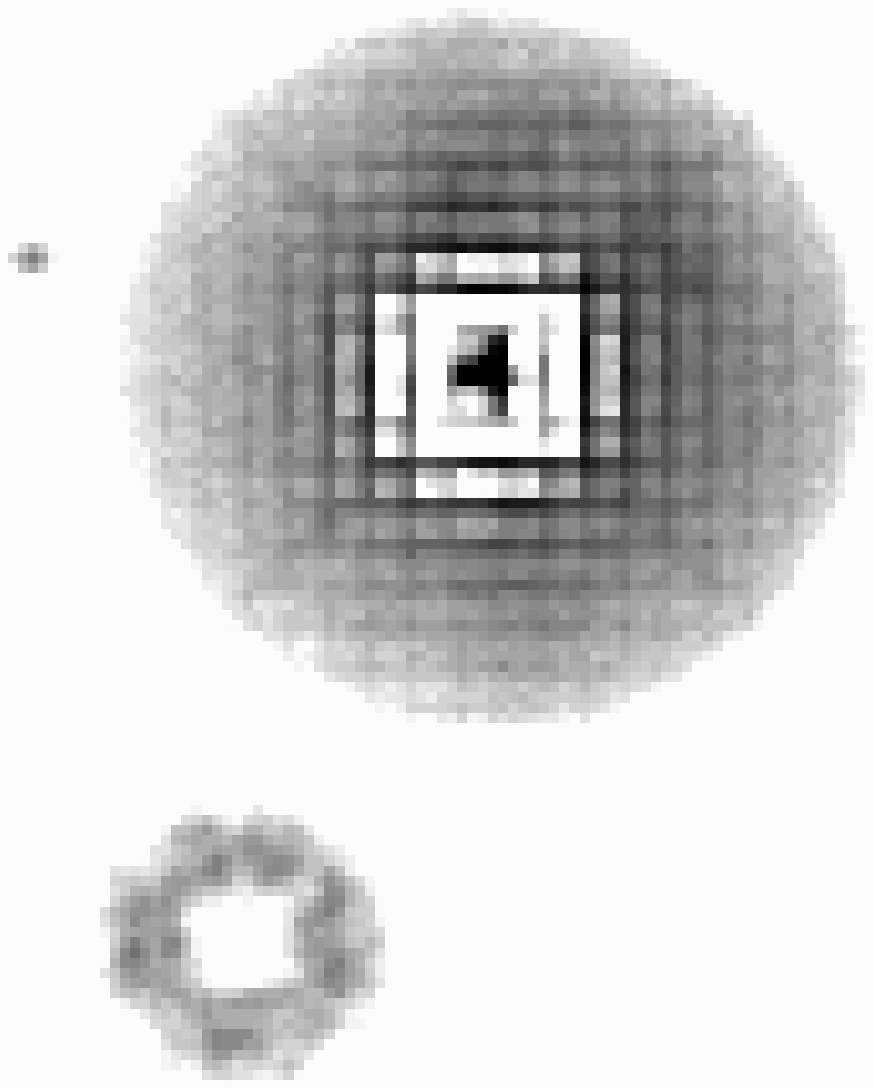}}
\end{picture}
\caption{The dynamic scale in these images has been chosen to enhance 
the straylight features.
{\it (a) (Top panel):} An out-of-focus ghost image of a bright star, 
taken from the 3C273 field. This star has a photographic magnitude of 10.7. 
{\it (b) (Bottom panel):} Straylight ellipses caused by reflection of a star outside
the FOV, taken from PKS 0312 offset 6 field. The average background 
count rate is 15\cpix; in the bright straylight loop it is 30\cpix.
The background is also enhanced in the central region due to reflection of 
diffuse sky light from outside the field. In the centre it rises to
$\sim 3$ times the background, in the V band.}
\label{figure8} 
\end{center}
\end{figure}

Artifacts can appear in XMM-OM images due to
light being scattered within the detector. These have two causes: internal
reflection of light within the detector window and reflection of off-axis
starlight and background light from part of the detector housing.

The first of these causes a faint, out of focus ghost
image of a bright star displaced in the radial direction away from the
primary image due the curvature of the detector window (Fig. 8a).

The second effect is due to light reflecting off a chamfer in the detector window housing.
Bright stars
that happen to fall in a narrow annulus 12.1 to 13 arc minutes off
axis shine on the reflective ring and form extended loops of emission
radiating from the centre of the detector (Fig. 8b). 
Similarly there is an enhanced
``ring'' of emission near the centre of the detector due to diffuse 
background light falling on the ring (Fig. 8b).  

The reflectivity of the ring, and of the detector window, reduces with
increasing photon energy. Therefore these features are
less prominent when using the UV filters.

\section{Conclusion}

The first stage of commissioning and calibrating XMM-OM has been completed.
The instrument is fulfilling its role of extending the spectral coverage of
XMM-Newton into the ultraviolet and optical band, allowing routine
observations of targets simultaneously with EPIC and RGS. Specifically, the instrument 
has successfully been demonstrated to provide wide field simultaneous imaging
with the X-ray camera, simultaneous timing studies, and boresight information
to arcsecond accuracy.
A number of
results illustrating the scientific potential of XMM-OM are contained within
this volume.

\begin{acknowledgements}

We would like to thank all the people who have contributed to the instrument;
in its building, testing and operation in orbit, as well as those who
have analysed the calibration data. The author list only contains a small fraction
of those people involved. XMM-OM was built by a consortium led by the 
Principal Investigator, Prof. K.O.Mason, and comprising, in the UK, 
the Mullard Space Science Laboratory and Department of Physics and
Astronomy, 
University College London; in the USA,
University of California Santa Barbara, Los Alamos National Laboratory, \& Sandia
National Laboratory; and in Belgium, the Centre 
Spatial Liege \& the University of Liege. 

JMV acknowledges support from
the SSTC-Belgium under contract P4/05 and by the PRODEX XMM-OM Project.
The U.S. investigators acknowledge support from NASA contract
NAS5-97119.
The UK contribution was supported by the PPARC.
 
Based on observations obtained with XMM-Newton, an ESA science 
    mission with instruments and contributions directly funded by 
    ESA Member States and the USA (NASA)

\end{acknowledgements}

\end{document}